\documentclass[prb,twocolumn,showpacs,superscriptaddress,preprintnumbers,amssymb]{revtex4}
\usepackage{graphicx}
\usepackage{dcolumn}
\usepackage{bm}
\usepackage{wasysym}
\input epsf

\input epsf

\begin{document}

\title{Dynamical models and the phase ordering kinetics of the $s=1$ spinor condensate}
\author{Subroto Mukerjee}
\affiliation{Department of Physics, University of California,
Berkeley, CA 94720} \affiliation{Materials Sciences Division,
Lawrence Berkeley National Laboratory, Berkeley, CA 94720}
\author{Cenke Xu}
\affiliation{Department of Physics, University of California,
Berkeley, CA 94720}
\author{J.~E.~Moore}
\affiliation{Department of Physics, University of California,
Berkeley, CA 94720} \affiliation{Materials Sciences Division,
Lawrence Berkeley National Laboratory, Berkeley, CA 94720}
\date{\today}
\begin{abstract}
The $s=1$ spinor Bose condensate at zero temperature supports
ferromagnetic and polar phases that combine magnetic and
superfluid ordering. We investigate the formation of magnetic
domains at finite temperature and magnetic field in two dimensions
in an optical trap. We study the general ground state phase
diagram of a spin-1 system and focus on a phase that has a
magnetic Ising order parameter and numerically determine the
nature of the finite temperature superfluid and magnetic phase
transitions. We then study three different dynamical models: model
A, which has no conserved quantities, model F, which has a
conserved second sound mode and the Gross-Pitaevskii (GP) equation
which has a conserved density and magnetization. We find the
dynamic critical exponent to be the same for models A and F
($z=2$) but different for GP ($z \approx 3$). Externally imposed
magnetization conservation in models A and F yields the value $z
\approx 3$, which demonstrates that the only conserved density
relevant to domain formation is the magnetization density.
\end{abstract}
 \maketitle

\section{Introduction}

The field of cold atomic gases has witnessed an explosion of
experimental and theoretical research in the last ten years. The
study of these systems has combined ideas from various disciplines
of physics such as atomic physics, condensed-matter physics,
optics etc. Cold atomic systems have provided a testing ground for
some of the most fundamental principles of collective quantum
behavior like Bose-Einstein Condensation. Of particular interest
is the study of spinor condensates, which are condensates of atoms
with non-zero spin and have been the focus of intense
experimental~\cite{gorlitz, wieman2, stamperkurn,
hadzibabicnature} and theoretical~\cite{ho, zhoudemler, sengstock,
lamacraft} studies in recent years. The spin degree of freedom
opens up the possibility of interesting collective magnetic
behavior in these systems in addition to the phenomenon of
Bose-Einstein condensation. It has already been demonstrated that
the presence of spin greatly modifies the nature of the condensate
and superfluid transition in spinor condensates compared to those
without spin~\cite{ho,mukerjee}.

Spinor condensates have over the last few years been realized in
both magnetic and optical traps. The latter are more interesting
from the point of view of spin ordering, since the spin degree of
freedom is not frozen out. The most widely studied atomic systems
are those of the spin-1 alkali atoms $^{23}$Na and $^{87}$Rb.
These systems differ from each other in the nature of the
effective two-body interaction, which is antiferromagnetic in the
former and ferromagnetic in the latter. The condensates with
antiferromagnetic interactions are also called polar. Recent
advances have made it possible to image ferromagnetic domains in
optical traps of $^{87}$Rb, making it possible to study the
interesting physics of domain formation in
them~\cite{stamperkurn}. This technique requires the application
of a magnetic field, an additional tunable parameter which makes
the phase diagram of these systems interesting. Moreover, these
atoms have also been trapped in two dimensional geometries, where
the physics of collective behavior is often more exotic than in
higher dimensions~\cite{hadzibabicnature,stamperkurn}. The
importance of this experiment for basic condensed matter physics
is twofold: it probes both our understanding of phase-ordering
kinetics at finite temperature (when observed at the longest
times) and, as the temperature is lowered or the observation time
is shortened, our understanding of dynamics across quantum phase
transitions.

In this paper we will investigate magnetic domain formation in
spin-1 systems at finite temperature and magnetic field. The main
purpose of this study is to compare and contrast various plausible
dynamical models with respect to coarsening of a magnetic order
parameter. The quantity of primary interest, will be the dynamic
critical exponent $z$ which determines the rate of domain
formation at large times: the domain size $L$ grows with time as
$L \sim t^z$. We will examine the general phase diagram of spin-1
condensates in the presence of a magnetic field in an optical trap
and comment on the broken symmetries of the various ordered
phases. We will then choose the phase that is most convenient to a
study of magnetic domain formation and elucidate the similarities
and differences between dynamic models, highlighting the
importance of different conservation laws in the dynamics. We will
compare our results with existing ones wherever possible.

A natural question is how the stochastic time-dependent
Ginzburg-Landau (TDGL) approach in this paper is related to
previous studies using deterministic equations of motion, such as
the Gross-Pitaevskii equation for the condensate, plus quantum
kinetic theory for excited
states~\cite{saitoueda2,lamacraft,saitoueda3}. The answer is that
the correct description depends on experimental parameters such as
the time scale of observation and the normal-state population. The
time scale at which stochastic processes resulting from
interaction with the normal cloud become important can be
increased by decreasing the temperature of the system. The initial
instability in a finite trap is likely to be described correctly
by the deterministic theories in the literature; coupling to the
many degrees of freedom in the normal cloud is irrelevant for the
immediate dynamics of the condensate. However, the longer times
accessed in current and future experiments are expected to be
described by the theory developed here.  In other words, the
universal dynamical properties in the sense of critical phenomena
are described by the theories presented here at any finite
temperature, as long as the system is observed for a sufficiently
long time.  We believe that current experiments may already be in
the regime where the theory presented here is valid.  However,
even if they are not, increases in observation time will soon
enable a precise comparison between theory and experiment.

Our main results on phase ordering of spinor condensates are
contained in sections VII and VIII. We argue in the final
discussion that one specific dynamical model (``model F''
dynamics, in the notation of the review paper of Hohenberg and
Halperin~\cite{hohenberghalperin}) is expected to describe the
long-time dynamics of spinor condensates.  This dynamical model is
a more complicated version of the model used in earlier studies of
superfluids~\cite{pu,robins,saitoueda,zhangzhou}, and reproduces
the known propagating modes of the spinor condensate at zero
temperature. All parameters in the dynamical model can be
determined from measurements of the condensate, as explained in
the appendix.

\section{The magnetic phase diagram of spin-1 bosons in an optical trap}

Spin-1 condensates are theoretically more complex than those with
zero spin~\cite{ho,leggettrev} in that the condensate order
parameter is a three component complex vector
\begin{equation}
\Psi = \left(\begin{array}{c} \psi_{+1} \\ \psi_{0} \\ \psi_{-1}
\end {array} \right),
\end{equation}
with $\psi_\alpha$ being the order parameter in the spin state of
eigenvalue $\alpha$ along some arbitrarily chosen direction. If
one assumes that the condensate state is a single particle zero
momentum state, the total energy for a given density of atoms in
an optical trap with a magnetic field $B$ in the $z$ direction can
be written as
\begin{equation}
E = c_2 \langle \vec{\bf S} \rangle^2 + g_2 \langle S_z^2 \rangle.
\end{equation}
Here $\vec{\bf S} = S_x \hat{x} + S_y \hat{y} + S_z \hat{z}$,
where
\begin{eqnarray} S_x & = & \frac{1}{\sqrt{2}}\left(
\begin{array}{ccc}
0 & 1 & 0 \\
1 & 0 & 1 \\
0 & 1 & 0 \end{array} \right) \\ \nonumber S_y & = &
\frac{i}{\sqrt{2}}\left( \begin{array}{ccc}
0 & -1 & 0 \\
1 & 0 & -1 \\
0 & 1 & 0 \end{array} \right) \\ \nonumber S_z & = & \left(
\begin{array}{ccc}
1 & 0 & 0 \\
0 & 0 & 0 \\
0 & 0 & -1 \end{array} \right)
\end{eqnarray}
are the generators of $SU(2)$ in the spin-1 representation and
$\langle A \rangle = \Psi^{\dagger}A\Psi$. $c_2$ is the spin-spin
interaction which can be antiferromagnetic ($c_2 > 0$) or
ferromagnetic ($c_2 < 0$). $g_2 \propto B^2$ and the second term
is just the quadratic Zeeman term. The absence of a linear term is
due to the fact that the time for the relaxation of magnetization
in optical traps is less than the lifetime of the condensate
itself. The ground state manifolds can be obtained by minimizing
the free energy with respect to $\{\psi^*_\alpha \}$. It has
already been shown that in the absence of a magnetic field, the
ground state manifolds in the polar and ferromagnetic cases are
isomorphic to the spaces $\frac{U(1) \times S^2}{Z_2}$ and $SO(3)$
respectively~\cite{mukerjee}. The phase diagram in the presence of
a magnetic field is given below.
\begin{figure}[h!]
\epsfxsize=3in \centerline{\epsfbox{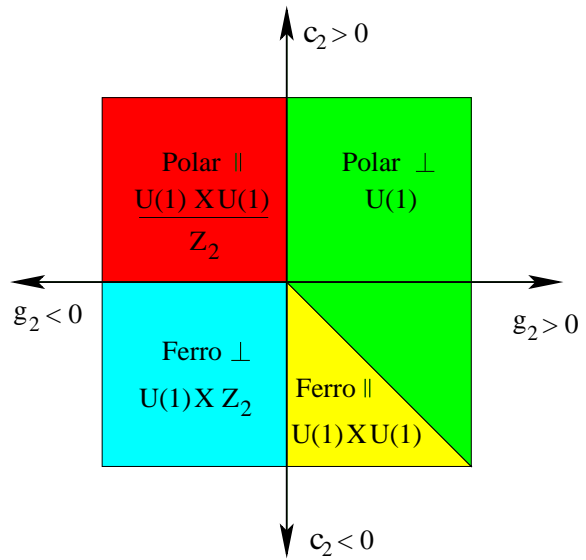}} \caption{The ground
state phase diagram of a spin-1 condensate in an optical trap in
the presence of a magnetic field that couples through a quadratic
Zeeman term. The different quadrants have different phases with
various types of in-plane and out-of-plane ordering. This figure
has been taken from Mukerjee {\em et. al}~\cite{mukerjee}.}
\label{phasediagram}
\end{figure}

Fig.~\ref{phasediagram} has four quadrants labelled by signs of
$c_2$ and $g_2$. In the polar ($c_2 > 0$) case, the magnetic
ordering is either in plane or out of plane depending on the sign
of $g_2$. ``Magnetic ordering'' here refers to the ordering of the
spin-quantization axis ($\hat{\bf n}$). The ground state is always
a macroscopically occupied single particle state of zero spin
projection in this case. For $g_2 > 0$, the only symmetry that is
broken in the ordered state is that of the $U(1)$ phase ($\theta$)
of the condensate. For $g_2 < 0$, however there is an additional
$U(1)$ due to the in-plane ordering of the spin-quantization axis.
The phase and spin are coupled through a $Z_2$ identification,
which denotes symmetry under $\theta \rightarrow \theta + \pi$ and
$\hat{\bf n} \rightarrow -\hat {\bf n}$. The vortices
corresponding to the spin and phase are thus coupled and can lead
to interesting finite-temperature physics in two
dimensions~\cite{podolsky}.

The lower part of the phase diagram corresponds to the
ferromagnetic case ($g_2 < 0$) and will be of primary interest to
us. The lower left quadrant corresponds to the case $g_2 < 0$. The
ground state now breaks a $U(1)$ symmetry corresponding to the
phase and an Ising $Z_2$ symmetry corresponding to the spin.
Physically this means that in the condensate, the bosons are
either in a state of spin projection 1 or -1. It is this Ising
degree of freedom, we exploit to study domain formation in two
dimensions. The reason is that since long range Ising order is
possible in two dimensions (as opposed to $U(1)$ order), it is
easier to define and measure the sizes of large magnetic domains
required to investigate long time behavior. It is thus this
quadrant that will be the focus of the rest of out studies. For
the sake of completeness we note that the lower right quadrant,
which corresponds to the case $g_2 > 0$ is divided into two parts
by a straight line with equation $g_2 = 2c_2$. To the left of this
line, one has in-plane ferromagnetic ordering with the spins
pointing in some $U(1)$ direction in the plane. The ground state
thus breaks two $U(1)$ symmetries, one corresponding to the phase
and the other corresponding to the spin. To the right of the line,
it is energetically favorable for the system to be in a polar out
of plane state. This suggests the interesting possibility of a
quantum phase transition in these systems tuned by the magnetic
field.

\section{2D finite temperature phase transitions}

Since we are interested in studying finite temperature coarsening
dynamics in the 2D system with $c_2 < 0$ and $g_2 < 0$, it is
important for us to locate the position of the superfluid and
magnetic transitions. This problem is also interesting in its own
right since such situations also come up in the study of classical
frustrated spin systems, like the fully frustrated $XY$
antiferromagnet (with $\pi$ flux per plaquette) on a square
lattice or the triangular lattice $XY$ antiferromagnet, where the
$Z_2$ corresponds to a chirality. The $U(1)$ and $Z_2$ transitions
are in close proximity to each other in these cases. The situation
is our particular case is not very different. We find that the
$U(1)$ transition is of the Kosterlitz-Thouless (KT) type and the
$Z_2$ transition of the 2D Ising type. Furthermore, we find that
for a certain range of parameters, $T_{Z_2} > T_{U(1)}$ which is
also what is observed in the fully frustrated $XY$ model on the
square lattice~\cite{olsson} and for others the order of the
transitions appears to be reversed. This depends on the magnitude
of the ratio of the parameters $g_2/c_2$. For small values of this
ratio, $T_{Z_2} > T_{U(1)}$. There is presumably also a point
where the two transitions occur at exactly the same temperature,
where the combined transition is in a different universality class
from 2D Ising and KT. We present here numerical data on just one
set of parameters where $T_{Z_2} > T_{KT}$ and illustrate how the
two transitions can be accurately determined despite being
reasonably close to each other in temperature. The method used is
due to Olsson~\cite{olsson} and we employ a numerical Monte-Carlo
simulations that uses the following Ginzburg-Landau free energy
functional

\begin{widetext}
\begin{equation}
F = \int d {\bf r} \left[\alpha \nabla \psi_a^* \nabla
\psi_a + a_0(T-T^{MF}_c) \psi_a^* \psi_a + \frac{c_0}{2}\psi_a^*
\psi_b^* \psi_b \psi_a + \frac{c_2}{2} \psi_a^* \psi_{a'}^* {\bf
S}_{ab}.{\bf S}_{a'b'}\psi_{b'}\psi_b + g_2 \psi^*_a \left(
S^2_z\right)_{ab}\psi_b\right]. \label{ginzburg1}
\end{equation}
\end{widetext}
with the following set of parameters, $\{\alpha=0.5, a_0=5.5,
c_0=7.0, c_2=-2.4, g_2=-1.3\}$.
\begin{figure}[h!]
\epsfxsize=4in \centerline{\epsfbox{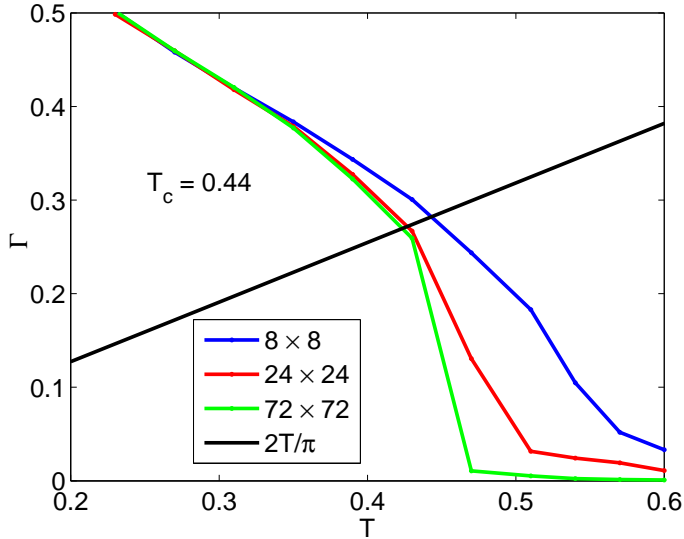}} \caption{The
helicity modulus as function of $\Gamma$ for the parameter set $\{
\}$ for different system sizes. A clear jump is visible of
$2T_c/\pi$ is visible at $T_c \approx 0.44$.} \label{KTtransition}
\end{figure}

The Kosterlitz-Thouless (KT) transition is detected by observing
the temperature dependence of the helicity modulus $Y$. The
helicity modulus for a discrete system of $N$ lattice points is
defined as
\begin{equation}
\Gamma = \left. \frac{1}{N}\frac{\partial^2<F>}{\partial \delta^2}
\right|_{\delta=0}
\end{equation}
where $\delta$ is a flux twist applied along a particular
direction. The helicity modulus undergoes a jump of magnitude
$\frac{2T_c}{\pi}$ at the location of the transition. This is
shown in Fig.~\ref{KTtransition}, where the transition temperature
is seen to be $T_c \approx 0.44$.

\begin{figure}[h!]
\epsfxsize=3in \centerline{\epsfbox{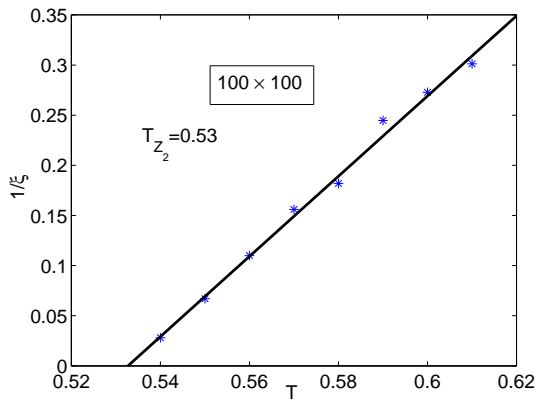}} \caption{The
magnetic correlation length as a function of temperature fitted to
the 2D Ising form. $T_c \approx 0.53$ as estimated this way.}
\label{Z2transition}
\end{figure}

The standard method to determine the location of the $Z_2$
transition using fourth order cumulants of the magnetic order
parameter fails here for the same reason that it does in the case
of the fully frustrated $XY$ model, which is the proximity to the
KT transition~\cite{olsson}. The cumulant method assumes that the
only relevant length scale at the transition is the system size
which is not true here because of the large correlation length
corresponding to the closely situated KT transition. Thus, a more
accurate method is to look at the critical exponent of the
correlation length of the magnetic order parameter and determine
$T_c$ by fitting it to the expected 2D Ising form. The
magnetization $M({\bf r})$ is given by
\begin{equation}
M({\bf r}) = |\psi_{+1}({\bf r})|^2-|\psi_{-1}({\bf r})|^2.
\end{equation}
The correlation length $\xi(T)$ can be extracted from the magnetic
autocorrelation function
\begin{equation}
g(r) = \langle M({\bf r})M(0) \rangle = e^{-r/\xi(T)}.
\end{equation}
If the transition is 2D Ising like,
\begin{equation}
\xi(T) \sim \frac{1}{T-T_c}.
\end{equation}
The numerical result is shown is Fig.~\ref{Z2transition}, which
shows that the correlation length fits the 2D Ising form fairly
well. The obtained transition temperature is $T_c \approx 0.53$. A
more careful finite-size scaling analysis can be done to determine
the two transition temperatures, but even at this level of
analysis it is clear that $T_{Z_2} > T_{U(1)}$.

\section{Dynamical models}

The study of the formation of domains of the order parameter
requires careful consideration of the dynamical modes of the
system. Dynamical models are often constrained by conservation
laws that are present as a consequence of symmetries or otherwise
in the system. It is well known that the presence of conservation
laws usually affects the rate of formation of domains, since the
phase space of states that the system can pass through in the
approach to the ordered state is constrained by the conservation
laws. However, not all conservation laws affect domain formation
in the same way and some might be more important than others. In
this section we consider some dynamical models appropriate for the
description of our system and comment on the conservation laws and
the dynamical modes obtained from them.

The most commonly used dynamical model to describe spinor
condensates is the Gross-Pitaevskii (GP)
equation~\cite{leggettrev}. This is the model that has been
extensively used to study domain formation in these systems. The
model consists of treating the condensate as a classical field at
zero temperature whose dynamics are given by the Hamilton
equations of motion of the appropriate Hamiltonian. In our case,
the Hamiltonian is
\begin{widetext}
\begin{equation}
H = \int d {\bf r} \left[\frac{\hbar^2}{2m} \nabla \psi_a^* \nabla
\psi_a + U({\bf r}) \psi_a^* \psi_a + \frac{c_0}{2}\psi_a^*
\psi_b^* \psi_b \psi_a + \frac{c_2}{2} \psi_a^* \psi_{a'}^* {\bf
S}_{ab}.{\bf S}_{a'b'}\psi_{b'}\psi_b + g_2 \psi^*_a \left(
S^2_z\right)_{ab}\psi_b \right],
\end{equation}
\end{widetext}
with the dynamical equation of motion
\begin{equation}
i\hbar\frac{\partial \psi_a}{\partial t} = -\frac{\delta H}{\delta
\psi^*_a}
\end{equation}
It should be noted that the condensate density $\psi^*_a\psi_a$
and magnetization are both conserved by this equation. However,
the GP equation cannot correctly describe the approach to
equilibrium at finite temperatures, since the dynamics is only
precessional and not relaxational. While this equation might be
appropriate for the description of the dynamics once the
condensate has been formed, it is inappropriate for the study of
dynamic phenomena in other cases, for example quenches from high
temperature where the energy of the condensate is not conserved.

The effect of finite temperature on the dynamics in spinor systems
has thus far been taken into account through phenomenological rate
equations, which too do not describe the approach to equilibrium.
A simple model which is more appropriate is the so-called ``model
A'' of the Hohenberg and Halperin
classification~\cite{hohenberghalperin}. This model uses the
Ginzburg-Landau free energy Eqn.~\ref{ginzburg1} with simple
Langevin dynamics. Operationally, this means the dynamical
equation
\begin{equation}
\frac{\partial \psi_a}{\partial t} = -\Gamma_0\frac{\delta
F}{\delta \psi^*_a} + \zeta_a({\bf r}, t). \label{modela}
\end{equation}
Thermal fluctuations due to finite temperature are contained in
the noise variable $\zeta_a({\bf r}, t)$, which has the following
autocorrelation function
\begin{equation}
\langle \zeta_a^*({\bf r}, t)\zeta_b({\bf r'}, t')\rangle =2{\rm
Re}(\Gamma_0) k_BT\delta({\bf r}-{\bf r'})\delta(t-t')\delta_{ab}
\end{equation}
consistent with the fluctuation-dissipation theorem, that drives
the system to equilibrium from a non-equilibrium state. This model
like the GP model only considers the condensate as a classical
field but unlike the GP model does not possess any conservation
laws. It is relaxational in nature with the rate of relaxation of
the order parameter set by Re $\Gamma_0$ and can be a complex
number. The condensate density is no longer conserved and neither
is the magnetization. The condensate is exchanging particles and
energy with the ``normal fluid'' in this model. The
non-conservation of energy of this model can be rectified by
implicitly including the normal fluid through a conserved ``second
sound'' mode ($m$), which is a real scalar field. Based on
previous experience with the superfluid transition in helium, one
expects that a correct description of the dynamics near the
transition requires this additional field and in the notation of
Hohenberg and Halperin, this model  ``model F'', with random
forces $\zeta_a$ and $\theta$ for the fields $\psi_a$ and $m$
respectively. The free energy $F_{\rm SS}$ with this second-sound
mode is
\begin{equation}
F_{\rm SS} = F + \int d{\bf r} \left(\gamma_0m\psi^*_a \psi_a +
\frac{1}{2C_0}m^2 \right), \label{ginzburg2}
\end{equation}
where $F$ is given by Eqn.~\ref{ginzburg1}. The dynamics are given
by
\begin{eqnarray} {\partial \psi \over \partial t} & = & -\Gamma_0
{\delta F_{\rm SS} \over \delta \psi^*_a} - i g_0 \psi {\delta
F_{\rm SS} \over \delta m} + \zeta_a({\bf r}, t) \\ \nonumber
{\partial m \over \partial t} & = & \lambda_0^m \nabla^2 {\delta
F_{\rm SS} \over \delta m} + 2 g_0 {\rm Im} \left(\psi^*_a {\delta
F_{\rm SS} \over \delta \psi^*_a}\right) + \tau({\bf r}, t)
\end{eqnarray}
These equations conserve the second sound density $m$ and the
noise correlator for $\theta$, consistent with the
fluctuation-dissipation theorem is
\begin{equation}
\langle\tau({\bf r},t)\tau({\bf r'},t')\rangle = -2\lambda_0^m
\nabla^2\delta({\bf r} - {\bf r'})\delta(t - t')
\end{equation}
The free energy $F_{\rm SS}$ has extra terms compared to $F$; a
term that couples $m$ and the condensate density and others that
contain the energy of the mode $m$. The dynamical equations also
incuding coupling terms as a consequence of the non-vanishing
Poisson brackets
$\{m,\psi_a\}$~\cite{halperinhohenberg,andersonsf,pitaevskii}.
Aside from terms that result from derivatives of the
Ginzburg-Landau free energy, there could in principle be
additional magnetic terms analogous to those in the Heisenberg
ferromagnet (e.g., ${\bf S} \times \nabla^2 {\bf S}$, where ${\bf
S}$ is the local spin density). Since this is even in ${\bf S}$,
it will not be obtained as the $S$ derivative of any free energy,
but will originate in the microscopic Hamiltonian.  A check that
no such additional terms are necessary is that the above equations
reproduce the previously obtained modes in the GP equation. Such a
calculation was carried out by Hohenberg and Halperin for the case
of Helium and we extend that to the case of spinor condensates in
the next section.

The three different dynamical models, GP, model A and model F are
the ones we will use to investigate domain formation in spinor
condensates at finite temperature and magnetic fields. We will in
addition to the dynamical equations above also impose the
conservation of magnetization on models A and F, to investigate
the effect of that conservation law on the dynamics. As can be
seen from the above discussion, model F contains many more
parameters than model A and the GP equation. The parameters of
this model are related to possible experimentally-measurable
quantities in the appendix.

\section{Dynamical modes in the ordered state}

Model F contains in it both the GP equation and model A, which can
be seen by setting the appropriate parameters in it to zero.
However, it is important that model F produces all the dynamical
modes that the GP equation does even when these parameters are not
zero, in order for this treatment to be valid. There will also be
additional modes produced (for example in $m$), that are absent in
the GP equation. We explicitly demonstrate this in this section.

We begin by setting the temperature and magnetic field to zero, to
enable comparison with the GP equation. The idea is to check that
the introduction of the extra parameters of model F does not alter
the modes that have already been calculated ~\cite{ho}. It is
known that there are three linearly dispersing mode in the polar
case and one gapped, linear and quadratic mode each in the
ferromagnetic case. The dynamical equations describing the modes
(either propagating or diffusion) are
\begin{eqnarray}
\frac{\partial\psi_\alpha}{\partial t} & = & -
\Gamma_0\frac{\delta F_{\rm SS} }{\delta \psi^\ast_\alpha} -
ig_0\psi\frac{\delta F_{\rm SS}}{\delta \psi_m}\\
\nonumber
\frac{\partial m}{\partial t} & = &
\lambda_0^m\nabla^2\frac{\partial F_{\rm SS} }{\partial m} + 2g_0
{\rm Im}(\psi^\ast_\alpha\frac{\delta F_{\rm SS}}{\delta
\psi^\ast_\alpha})\label{eq0}
\end{eqnarray}
For brevity of notation, we also set $T = -a \nabla^2$,
$\mu=a_0T^{MF}_c$ and explicitly write $\Gamma_0 = \Gamma_1 + i
\Gamma_2$, where both $\Gamma_1$ and $\Gamma_2$ are real.

\subsection{The polar case}
We assume that
\begin{equation}
\Psi = N + \Phi,
\end{equation}
where
\begin{equation}
N = \sqrt{n_0}\left(
\begin{array}{cc}
0  \\
1  \\
0  \end{array} \right),
\end{equation}
is the value of the order parameter in the ordered polar state and
$\Phi$ is a perturbation on it. The number of particles within the
condensate, is related to the value $\mu$ and $c_0$ by minimizing
the free energy $F_{\rm SS}$
\begin{equation}
\sqrt{n_0} = \sqrt{\frac{\mu}{c_0}} \label{n0}
\end{equation}

\subsubsection{Polar state with $\Gamma_1$ = 0, $\lambda_0 = 0$ ,
$\gamma$ = 0}

Let us first ignore the coupling $\gamma m
\psi^\ast_\alpha\psi_\alpha$, as well as all the dissipation terms
in the equations, like the term with coefficient $\Gamma_1$ and
$\lambda_0$. We will put them back in later.

In this case, all the modes are propagating, since there is no
dissipation. After expanding the equations around $N$, the
linearized equations we have are
\begin{eqnarray}\nonumber
\frac{\partial\phi_0}{\partial t} & = & i\Gamma_2(T\phi_0 +
2c_0n_0(\phi_0 + \phi_0^\ast)) + im\frac{\sqrt{n_0}}{C}g_0 \\
\nonumber
 \frac{\partial m}{\partial t} & = &
-ig_0\sqrt{n_0}\nabla^2(\phi_0 - \phi_0^\ast)\\
\nonumber
\frac{\partial \phi_1}{\partial t} & = &
-i\Gamma_2(T\phi_1 + n_0c_2(\phi_1 + \phi^\ast_{-1}))\\
\frac{\partial \phi_{-1}}{\partial t} & = & -i\Gamma_2(T\phi_{-1}
+ n_0c_2(\phi^\ast_1 + \phi_{-1}))\label{eq1}
\end{eqnarray}
Notice that in order to get these equations, we need $\mu$ to take
exactly the value in (\ref{n0}))

The equations for $\phi_1$ and $\phi_{-1}$ have the same form as
for the GP equation~\cite{ho}, so the spin wave modes are the
same. Both $M_+= \phi_{+1}+i\phi_{-1}$ and
$M_-=\phi_{+1}-i\phi_{-1}$ disperse linearly with $k$, with
velocity $c_M = \Gamma_2\sqrt{n_0c_2}$.

The step by step solution for the coupled equation between
$\phi_0$, $\phi^\ast_0$ and $m$ is tedious, so we only write down
the result here. Basically, the second sound mode and density
fluctuation $\delta n_0 = \phi^\ast_0 + \phi_0$ couple and form
two modes with linear dispersion relations, the velocity is
\begin{equation}
c_s = \sqrt{\frac{g^2_0n_0}{C} + 2a\Gamma_2^2c_0n_0}\label{cs}
\end{equation}

Notice that the first term in the square root in the above
equation is the square of the second sound
velocity~\cite{hohenberghalperin}, with $\Gamma_2 = 0$ and
ignoring the propagating mode of $\psi$. The second term in the
square root is the one that appears as the density fluctuation
mode in Ref. 5, where the the second sound mode was ignored. Here
we see that if we take into account both densities, the second
sound mode and the density fluctuation mode couple into a new mode
with velocity $c_s$.

\subsubsection{Polar state with $\Gamma_1 = 0$,  $\lambda_0
\neq 0$, $\gamma = 0$}

Here the dissipation $\lambda_0$ term is added back into the
equations. We will not consider the case with finite $\Gamma_1$,
since we assume that within the condensate, the dissipation of the
modes is very small.

The coupled equations between $m$ and $\phi_0$ now become
\begin{eqnarray}
\frac{\partial\phi_0}{\partial t} & = & i\Gamma_2(T\phi_0 +
2c_0n_0(\phi_0 + \phi_0^\ast)) + im\frac{\sqrt{n_0}}{C}g_0 \\
\frac{\partial m}{\partial t} & = & -ig_0\sqrt{n_0}\nabla^2(\phi_0
- \phi_0^\ast) + \frac{\lambda_0}{C}\nabla^2m \label{eq2}
\end{eqnarray}

The detailed solution is again tedious and we will solve the
equation based on following approximation that the higher order
terms of spatial derivatives are small, since we are only
interested in the limit $k$ going to zero. Under this
approximation the dispersion relation is
\begin{equation}
\omega = c_sk + i\frac{\lambda_0}{C}k^2
\end{equation}

The mode gets a propagating part, which is linear in $k$, and a
damping part, which is proportional to $k^2$. $c_s$ is given by
(\ref{cs}).

\subsubsection{Polar state with $\Gamma_1 = 0$,  $\lambda_0 \neq
0$, $\gamma \neq 0$}

Turning on $\gamma$, changes only two terms in the equations.
First,
\begin{equation}
\frac{\partial \phi_0}{\partial t} = i\Gamma_2(T\phi_0 +
2c_0n_0(\phi_0 + \phi_0^\ast)) + ig_0\gamma n_0(\phi_0 +
\phi_0^\ast) + im\frac{\sqrt{n_0}}{C}g_0.
\end{equation}
We can redefine
\begin{equation}
c_0^\prime = c_0 + \frac{g_0\gamma}{\Gamma_2}
\end{equation}
and make the equation look exactly like (\ref{eq1}), except for
replacing $c_0$ by $c_0^\prime$. $\gamma$ also modifies
(\ref{eq2}), by adding a term to the right hand side
\begin{eqnarray}\nonumber
\frac{\partial\phi_0}{\partial t} & = & i\Gamma_2(T\phi_0 +
2c_0n_0(\phi_0 + \phi_0^\ast)) + im\frac{\sqrt{n_0}}{C}g_0 \\
\nonumber \frac{\partial m}{\partial t} & = &
-ig_0\sqrt{n_0}\nabla^2(\phi_0 - \phi_0^\ast) +
\frac{\lambda_0}{C}\nabla^2m \\ & & +
\frac{\lambda_0\gamma}{C}\sqrt{n_0}\nabla^2(\phi_0 + \phi_0^\ast)
\end{eqnarray}
Solving this modified equation, we see that the term proportional
to $\gamma$ only contributes higher order momentum terms. So, up
to linear order in $k$, the dispersion relation is not changed.
Therefore all the results here are the same as case 2, if we
replace $c_0$ by $c_0^\prime$.

\subsection{The ferromagnetic case}

The solution of the ferromagnetic case is very similar to the
polar case. The general formalism and effective action
Eqn.~\ref{ginzburg2} still apply. The difference is in how we
linearize the equations. In the ferromagnetic case, we should
linearize the equations around the state
\begin{equation}
\Psi = N + \Phi,
\end{equation}
with
\begin{equation}
N = \sqrt{n_0}\left(
\begin{array}{cc}
1  \\
0  \\
0  \end{array} \right)
\end{equation}
Here the density of the condensate is not only related to the
coefficient $c_0$, but also to the coefficient $c_2$.
\begin{equation}
\sqrt{n_0} = \sqrt{\frac{\mu}{c_0+c_2}} \label{n02}
\end{equation}
We now obtain the following linearized equations
\begin{eqnarray} \nonumber
\frac{\partial\phi_1}{\partial t} & = & i\Gamma_2(T\phi_0 + 2(c_0
+ c_2)n_0(\phi_1 + \phi_1^\ast)) + im\frac{\sqrt{n_0}}{C}g_0 \\
\nonumber
 \frac{\partial m}{\partial t} & = &
-ig_0\sqrt{n_0}\nabla^2(\phi_1 - \phi_1^\ast) +
\frac{\lambda_0}{C}\nabla^2m \\ \nonumber
\frac{\partial\phi_0}{\partial t} & = & -i\Gamma_2T\phi_0\\
\frac{\partial \phi_{-1}}{\partial t} & = & -i\Gamma_2T\phi_{-1} +
2c_2n_0\phi_{-1}\label{eq3}
\end{eqnarray}
The dispersion relation for the second sound mode is
\begin{eqnarray}
\omega & = & c_sk + i\frac{\lambda_0}{C}k^2 \\\nonumber c_s & = &
\sqrt{\frac{g^2_0n_0}{C} + 2a\Gamma_2^2(c_0 + c_2)n_0},\label{cs2}
\end{eqnarray}
where we have kept terms only to lowest order in the momentum in
every step of the calcultion. For the density fluctuation mode
$\delta n = \sqrt{n_0}(\phi_1 + \phi_{1^\ast})$, the dispersion
relation is
\begin{equation}
\omega = c_sk
\end{equation}
The spin wave mode $\delta M_- = \sqrt{n_0}\phi^\ast_0$ has the
same dispersion relation as the one obtained in the GP
case~\cite{ho}, $\omega = ak^2$. Again, turning on the interaction
$\gamma$ does not change the result much. It causes a redefinition
of $c_0$ in the propagating part of the mode, and only contributes
higher order momentum terms in the diffusion or damping part of
the mode, which are not important when the momentum is small.

Thus, we see that in both the polar and ferromagnetic cases, the
dynamic modes obtained in the presence of the extra parameters of
model F are consistent with those obtained from the GP equations.
The nature of the density mode changes because of coupling with
the second sound mode but the spin wave modes remain unaffected.

\section{Domain formation}
A typical experiment or numerical simulation of coarsening
involves starting the system off at a high-temperature (usually
disordered) state and rapidly quenching it to a temperature below
the ordering transition to observe the growth of domains of the
ordered state. At the heart of the theoretical analysis of this
process is the scaling hypothesis~\cite{bray}. The equal-time
correlation function of the order parameter $m({\bf r}, t)$ is
defined as
\begin{equation}
C({\bf r}, t) = \langle m({\bf x}+{\bf r}, t)m({\bf r}, t)\rangle.
\end{equation}
The scaling hypothesis states that
\begin{equation}
C({\bf r}, t) = f\left( \frac{r}{L(t)}\right)
\end{equation}
$L(t)$ is a characteristic length scale, the domain size. Further,
at long times $L \propto t^{1/z}$, where $z$, the dynamical
critical exponent. The dynamical critical exponent is dependent on
the model used to describe the ordering dynamics of the system,
the symmetry of the order parameter and the nature of defects
present in the initial state. For instance, it is known that $z=2$
for the Ising model with dynamics which do not conserve the total
magnetization, after a high temperature quench. For an $XY$ model
on the other hand $z=2$ but with a logarithmic correction as a
function of time~\cite{bray}. The difference from the Ising case
can be attributed to the different broken symmetry and
consequently the topological defects present in the high
temperature state. The Ising model with a conserved order
parameter on the other hand produces a different dynamic critical
exponent $z=3$~\cite{bray,husedyn}. The growth of domains is
slower in this case compared to the case with no magnetization
conservation since the conservation law places constraints on the
phase space available during domain growth.

\section{Details of the Numerical Simulation}

In this work we study the dynamics of domain growth in 2D after a
magnetic field quench and not a temperature quench. The motivation
is a similar approach adopted in recent experiments in optical
traps. To be specific, we study a ferromagnetic condensate in two
dimensions whose initial state is a polar out-of-plane state in
the fourth quadrant of Fig.~\ref{phasediagram} and quench it to a
value of the field, where the ordered state is a ferromagnetic
out-of-plane state (in quadrant 3 of Fig.~\ref{phasediagram}).
Operationally, we sweep the parameter $g_2$ from a large positive
value to a negative value. For models A and F, this is done at
finite temperature and the order parameter eventually relaxes to a
uniform value consistent with the ferromagnetic out-of-plane
state. The Gross-Pitaevskii equation on the other hand does not
cause the system to relax but rather to oscillate between
different concentrations of the three spinor components. Further,
it does not allow an initial state which is completely polar
out-of-plane to form magnetic domains of the +1 and -1 components
at any value of the time. In this case, we start with an initial
state, which has 90$\%$ of the atoms in the 0 (polar out-of-plane)
state and the other 10$\%$, divided equally among the +1 and -1
states. The phase of each spinor component in the initial state is
chosen to be a random number between $0$ and $2\pi$ allowing for spatial
inhomogeneity which leads to domain formation.

The equations of motion corresponding to each model are integrated
numerically using a first order Euler method with the noise
functions drawn from a Gaussian distribution. The size of the
numerical grid ranges from $50 \times 50$ to $200 \times 200$. The
time step is adjusted depending on the values of the other
parameters and varied to check for consistency. The number of
parameters is large (especially for model F) and we present
results only for a fixed set of parameters. However, we have
explored other parts of the parameter space consistent with
ferromagnetic out-of-plane order and not found any qualitative and
wherever appropriate (like for the value of $z$) quantitative
difference in the results. The set of parameters for which we
report results are those in section III with the additional model
F parameters, $\{{\rm Re}(\Gamma_0)=1.30, {\rm Im}(\Gamma_0)=0.26,
g_0=0.35, \lambda_0^m=0.84, \gamma_0=1.5\}$, wherever applicable.

The domain size $L(t)$ is measured using the relation
\begin{equation}
L(t) = \sqrt{\frac{S_0(t)}{S_2(t)}}, \label{domcalc}
\end{equation}
where $S_0(t)$ and $S_2(t)$ are respectively the zeroth and second
moment of the structure function
\begin{equation}
S({\bf k}, t) = \int d{\bf r} \langle M({\bf r},t) M(0,t) \rangle
e^{i{\bf k}.{\bf r}},
\end{equation}
which is the Fourier transform of the order parameter correlation
function. The domain size is also calculated by measuring the size
of domain boundaries directly in the simulation grid. This method
serves as a consistency check on the first method. It should be
mentioned though that the second method is useful and consistent
with the first one only when there are very few small bubbles of
one value of the order parameter inside large islands of the other
value. This method essentially ignores these bubbles by looking
for large closed domain walls and works best when the domains are
large in size.

\section{Results}

\subsection{Finite temperature without conservation of
magnetization}

\begin{figure}[h!]
\epsfxsize=4in \centerline{\epsfbox{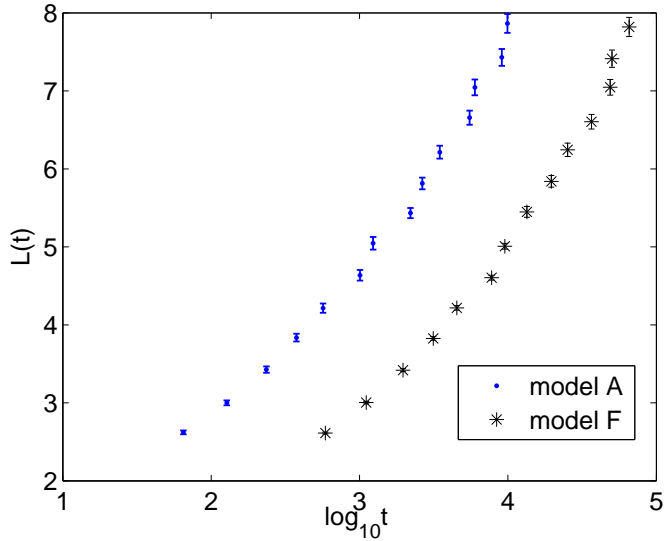}} \caption{$L(t)$ as
a function of $\log_{10}t$ for model F with the parameter set
${\mathcal R}$, with no magnetization conservation
\label{nomagd}.}
\end{figure}

\begin{figure}[h!]
\epsfxsize=4in \centerline{\epsfbox{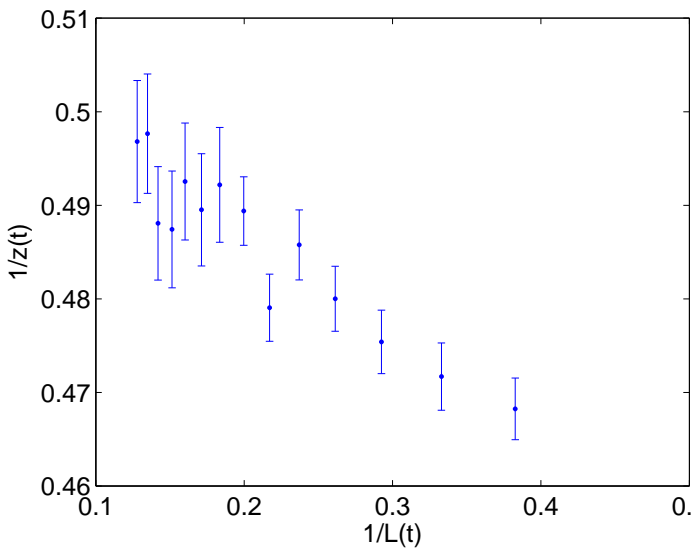}}
\caption{$1/z(t)$ as a function of $1/L(t)$ for model F and no
magnetization conservation with the parameter set ${\mathcal R}$
demonstrating the drift of $z(t)$ as a function of $t$ and thus
increasing $L(t)$ towards the value $z=2$.} \label{nomagdrift}
\end{figure}

We present results for the domain size as a function of time for
models A and F in Fig~\ref{nomagd}. The results presented are for
the set of parameters mentioned in the preceding section, with a
magnetic field quench and have been obtained on a grid of size
$200 \times 200$. It can be seen that domain formation is faster
for model A than for model F, which can be attributed to the
presence of the extra conservation law. This certainly appears to
be the case over the range of parameters that we have explored,
but may not be the case elsewhere in parameter space. Whether or
not this is a universal feature requires more careful analysis.
The curve for $L(t)$ as a function of $t$ for model A dynamics
seems to yield a $z = 2 \pm 0.15$ over the entire range of values
of time we have presented. Further, the value of $z$ obtained at
different values of time seems to be fairly constant. This is the
value of $z$, one would expect for a high temperature quench in a
pure Ising model. Model F also yields $z \approx 2$. Unlike in
model A dynamics, there is a small drift in the value of $z$
obtained at different values of $t$. A similar drift (of a larger
magnitude) has been seen in the case of the Ising model with
dynamics that conserve magnetization and it has been argued by
Huse~\cite{husedyn} that this is due to excess transport in domain
interfaces. It then follows that the effective dynamic critical
exponent $z(t)$ drifts in the following way to first order in the
domain size
\begin{equation}
\frac{1}{z(t)}=\frac{1}{z(t=\infty)}\left[ 1-\frac{L_0}{L(t)}
\right] \label{domdrift}
\end{equation}
This suggests that $z(t)$ approaches its infinite time value from
above, which appears to be the case here as well as can be seen
from Fig.~\ref{nomagdrift} , which is a plot of $1/z(t)$ vs.
$1/L(t)$. However, we emphasize that the above analysis is
strictly applicable only to the case where the {\em order
parameter is conserved}, which is not the case here. The quantity
that is conserved is the second sound mode. Nevertheless, it is
possible that the drift can be explained by some mechanism similar
to the above.

To conclude this part, we remark that both models A and F without
any explicit magnetization conservation both yield the same
dynamic critical exponent $z = 2$ for coarsening with a magnetic
field quench.

\subsection{Finite temperature with conserved magnetization}

\begin{figure}[h!]
\epsfxsize=4in \centerline{\epsfbox{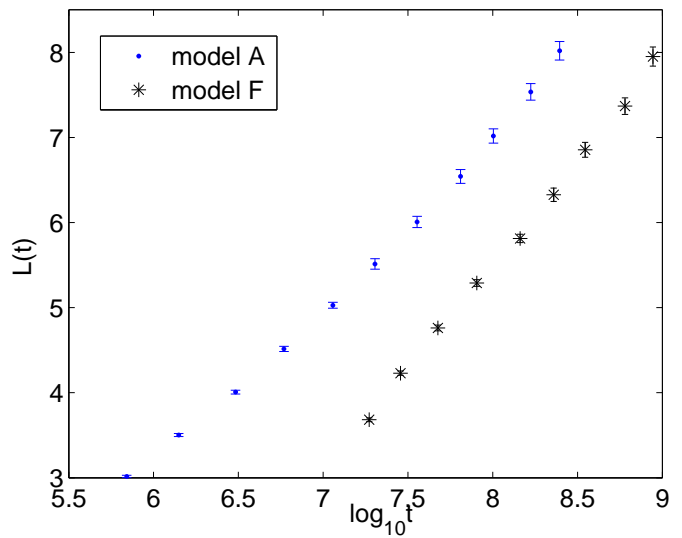}} \caption{$L(t)$ as a
function of $\log_{10}t$ for models A and F and conserved
magnetization density with the parameter set ${\mathcal R}$.}
\label{magd}
\end{figure}

\begin{figure}[h!]
\epsfxsize=4in \centerline{\epsfbox{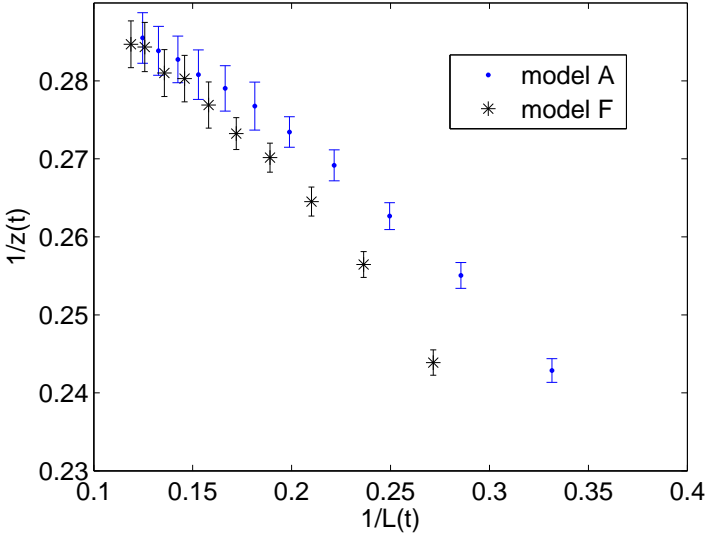}} \caption{$1/z(t)$
as a function of $1/L(t)$ for models A and F and conserved
magnetization density with the parameter set ${\mathcal R}$
demonstrating the drift of $z(t)$ as a function of $t$ and thus
increasing $L(t)$ towards the value $z=3$.} \label{magdrift}
\end{figure}

We now present results for models A and F with conserved
magnetization. The magnetization conservation is implemented in
terms of a local continuity equation in the magnetization density
and a magnetization current. We illustrate how we do this for
model A and the implementation for model F proceeds along similar
lines. We first note that the magnetization density
$M=|\psi_{+1}({\bf r},t)|^2-|\psi_{-1}({\bf r}, t)|^2$ only
involves the amplitudes of the componenets of the condensate order
parameter. We first write down model A dynamics in terms of
separate dynamical equations for the phase and amplitude of each
component of the condensate order parameter. These turn out to be
\begin{eqnarray}\nonumber
\frac{\partial |\psi_a|}{\partial t} & = & -\frac{1}{2}{\rm
Re}(\Gamma_0) \frac{\delta F}{\delta |\psi_a|} +
\frac{1}{2|\psi_a|}{\rm Im}(\Gamma_0)\frac{\delta F}{\delta
\theta_a} + \mu_a({\bf r}, t), \label{modelaconden}\\ & &
\\\nonumber
|\psi_a|\frac{\partial \theta_a}{\partial t} & = &
-\frac{1}{2|\psi_a|}{\rm Re}(\Gamma_0)\frac{\delta F}{\delta
\theta_a}+ \frac{1}{2}{\rm Im} (\Gamma_0)\frac{\delta F}{\delta
|\psi_a|} + \nu_a({\bf r}, t), \label{modelaconphase} \\ & &
\end{eqnarray}
where the noise correlators are
\begin{eqnarray}
\langle \mu_a({\bf r}, t)\mu_b({\bf r'}, t')\rangle &=&\langle
\nu_a({\bf r}, t)\nu_b({\bf r'}, t')\rangle \\\nonumber &=&{\rm
Re}(\Gamma_0) k_BT\delta({\bf r}-{\bf r'})\delta(t-t')\delta_{ab}.
\end{eqnarray}
Note that every quantity in the above equations is now real. If we
were interested in conserving the density $|\psi_a|^2$ of each
component {\em individually}, we would modify
Eqn.~\ref{modelaconden} to
\begin{eqnarray}\nonumber
\frac{\partial |\psi_a|^2}{\partial t} & = & -{\rm Re}(\Gamma_0)
\nabla^2\left( |\psi_a|\frac{\delta F}{\delta |\psi_a|}\right) +
{\rm Im}(\Gamma_0)\nabla^2\left( \frac{\delta F}{\delta
\theta_a}\right) \\ & & + \mu_a({\bf r}, t), \label{eachcomp}
\end{eqnarray}
with the correlator for $\mu$ now given by
\begin{equation}
\langle\mu_a({\bf r},t)\mu_a({\bf r'},t')\rangle = 4{\rm Re}
(\Gamma_0)\nabla^2[|\psi_a({\bf r},t)|^2\delta({\bf r} - {\bf
r'})]\delta(t - t').
\end{equation}
This ensures there is a conservation equation of the sort
\begin{equation}
\frac{\partial |\psi_a({\bf r},t)|^2}{\partial t} = -{\bf \nabla
.}{\bf J_{a}}({\bf r}, t)
\end{equation}
for each component. We are however not interested in conserving
the density of each component, but only the combination
$M=|\psi_{+1}|^2-|\psi_{-1}|^2$. To this end, proceeding as above,
we obtain the following set of equations.
\begin{widetext}
\begin{eqnarray}
\frac{\partial M}{\partial t} & = & -{\rm Re}(\Gamma_0)
\nabla^2\left( |\psi_{+1}|\frac{\delta F}{\delta |\psi_{+1}|} -
|\psi_{-1}|\frac{\delta F}{\delta |\psi_{-1}|}\right) + {\rm
Im}(\Gamma_0)\nabla^2\left(\frac{\delta F}{\delta
\theta_{+1}}-\frac{\delta F}{\delta \theta_{-1}}\right) +
\mu_M({\bf r}, t)\\
\frac{\partial N}{\partial t} & = & -{\rm Re}(\Gamma_0)\left(
|\psi_{+1}|\frac{\delta F}{\delta |\psi_{+1}|} +
|\psi_{-1}|\frac{\delta F}{\delta |\psi_{-1}|}\right) + {\rm
Im}(\Gamma_0)\left(\frac{\delta F}{\delta \theta_{+1}}+
\frac{\delta F}{\delta \theta_{-1}}\right) + \mu_N({\bf r}, t)\\
\frac{\partial |\psi_0|}{\partial t} & = & -\frac{1}{2}{\rm
Re}(\Gamma_0) \frac{\delta F}{\delta |\psi_0|} +
\frac{1}{2|\psi_0|}{\rm Im}(\Gamma_0)\frac{\delta F}{\delta
\theta_0} + \mu_0({\bf r}, t)\\
|\psi_a|\frac{\partial \theta_a}{\partial t} & = &
-\frac{1}{2|\psi_a|}{\rm Re}(\Gamma_0)\frac{\delta F}{\delta
\theta_a}+ \frac{1}{2}{\rm Im} (\Gamma_0)\frac{\delta F}{\delta
|\psi_a|} + \nu_a({\bf r}, t)
\end{eqnarray}
Here $N=|\psi_{+1}|^2+|\psi_{-1}|^2$ and the noise correlators are
given by
\begin{eqnarray}\nonumber
\langle\mu_M({\bf r},t)\mu_M({\bf r'},t')\rangle & = & 4{\rm Re}
(\Gamma_0)\nabla^2[\{|\psi_{+1}({\bf r},t)|^2+|\psi_{+1}({\bf
r},t)|^2\}\delta({\bf r} - {\bf r'})]\delta(t - t')\\\nonumber
\langle\mu_N({\bf r},t)\mu_N({\bf r'},t')\rangle & = & 4{\rm Re}
(\Gamma_0)\{|\psi_{+1}({\bf r},t)|^2+|\psi_{+1}({\bf
r},t)|^2\}\delta({\bf r} - {\bf r'})\delta(t - t')\\\nonumber
\langle \mu_0({\bf r}, t)\mu_0({\bf r'}, t')\rangle& = &{\rm
Re}(\Gamma_0)k_BT\delta({\bf r}-{\bf r'})\delta(t-t')\\
\langle \nu_a({\bf r}, t)\nu_b({\bf r'}, t')\rangle& = &{\rm
Re}(\Gamma_0) k_BT\delta({\bf r}-{\bf r'})\delta(t-t')\delta_{ab}
\end{eqnarray}
\end{widetext}
The noise functions $\mu_N$, $\mu_M$, $\mu_0$ and $\nu_a$ are
mutually uncorrelated. The above equations for $M$ and $N$ can be
used to generate equations for $|\psi_{+1}|$ and $|\psi_{-1}|$,
which is the way the numerical calculation is performed. Note that
the full set of dynamical equations written above has no
conservation law except the one for $M$. We now use the same
procedure to impose magnetization conservation on model F. The
dynamical equations in this case are
\begin{widetext}
\begin{eqnarray}
\frac{\partial M}{\partial t} & = & -{\rm Re}(\Gamma_0)
\nabla^2\left( |\psi_{+1}|\frac{\delta F_{ss}}{\delta |\psi_{+1}|}
- |\psi_{-1}|\frac{\delta F_{ss}}{\delta |\psi_{-1}|}\right) +
{\rm Im}(\Gamma_0)\nabla^2\left(\frac{\delta F_{ss}}{\delta
\theta_{+1}}-\frac{\delta F_{ss}}{\delta \theta_{-1}}\right) +
\mu_M({\bf r}, t)\\
\frac{\partial N}{\partial t} & = & -{\rm Re}(\Gamma_0)\left(
|\psi_{+1}|\frac{\delta F_{ss}}{\delta |\psi_{+1}|} +
|\psi_{-1}|\frac{\delta F_{ss}}{\delta |\psi_{-1}|}\right) + {\rm
Im}(\Gamma_0)\left(\frac{\delta F_{ss}}{\delta \theta_{+1}}+
\frac{\delta F_{ss}}{\delta \theta_{-1}}\right) + \mu_N({\bf r}, t)\\
\frac{\partial |\psi_0|}{\partial t} & = & -\frac{1}{2}{\rm
Re}(\Gamma_0) \frac{\delta F_{ss}}{\delta |\psi_0|} +
\frac{1}{2|\psi_0|}{\rm Im}(\Gamma_0)\frac{\delta F_{ss}}{\delta
\theta_0} + \mu_0({\bf r}, t)\\
|\psi_a|\frac{\partial \theta_a}{\partial t} & = &
-\frac{1}{2|\psi_a|}{\rm Re}(\Gamma_0)\frac{\delta F_{ss}}{\delta
\theta_a}+ \frac{1}{2}{\rm Im} (\Gamma_0)\frac{\delta
F_{ss}}{\delta |\psi_a|}-g_0|\psi_a|\frac{\partial
F_{ss}}{\partial m}+ \nu_a({\bf r}, t)\\
{\partial m \over \partial t} & = & \lambda_0^m \nabla^2 {\delta
F_{\rm SS} \over \delta m} + g_0 {\rm Im} \left(|\psi_a| {\delta
F_{\rm SS} \over \delta |\psi_a|}\right) + \tau({\bf r}, t)
\end{eqnarray}
\end{widetext}
The noise correlators are the same as for model A with
magnetization conservation with the additional correlator
\begin{equation}
\langle\tau({\bf r},t)\tau({\bf r'},t')\rangle = -2\lambda_0^m
\nabla^2\delta({\bf r} - {\bf r'})\delta(t - t'),
\end{equation}
as in the case of model F without magnetization conservation. It
should be noted that the coefficient $g_0$ appears only in the
dynamical equation for the phases of the different components of
the condensate thus making its identification as a precessional
term obvious. Further, the above dynamical equations conserve both
the magnetization $M$ and the second sound mode $m$ and thus
represent perhaps the most realistic dynamical model for a BEC at
finite temperature and field; one where the condensate can
exchange charge and energy with the ``normal cloud'' but not
magnetization.

Once again, the results presented are for a $200 \times 200$
simulation grid. We have checked that the additional magnetization
conservation law does not affect the static properties of the
model. As in the previous case, we again see that domain formation
is faster for model A and than model F. This time, however, the
dynamic critical exponent obtained is not equal to 2. As can be
seen from Fig.~\ref{magdrift}, which is a plot of $1/z(t)$ vs.
$1/L(t)$ for both models, there is a significant drift of $z(t)$
as a function of time. Once again the direction of the drift is
consistent with Eqn.~\ref{domdrift} and in this case, the analysis
mentioned in the previous subsection is directly applicable, since
it is the order parameter (the magnetization) that is directly
conserved. However, the noise levels of the simulations do not
permit a fit to Eq.~\ref{domdrift}. It should be noted though that
to the extent observable in the numerical simulation, there is a
drift in the direction of $z=3$ in the data. This is what is
observed in a pure Ising model with conserved magnetization in a
high temperature quench. One interesting observation is that the
value of $1/z(t)$ seems to deviate more strongly from
Eq.~\ref{domdrift} at large times for model F than model A. This
deviation has also been observed for the simple Ising model with
conserved magnetization, where it was attributed to finite-size
effects and correlated noise in the simulations. That could well
be the case here as well, although it is not clear why these
effects should be more pronounced in one model than in the other.
It should be noted that in the simulations on the Ising
model~\cite{husedyn}, the values of $z(t)$ observed for comparable
simulation times are roughly close to what we observe.

The conclusion of this part is that models A and F with explicit
magnetization conservation yield a dynamic critical exponent $z
\approx 3$, different from the exponent obtained without
magnetization conservation. Thus, the models A and F seem to be
identical as far as long-time coarsening behavior of the
magnetization is concerned and the behavior is truly determined by
whether or not the magnetization is conserved, which it is not
explicitly in either model. The difference between these two
models will become apparent, when the coarsening of domains
related to the conserved second mode is investigated.

\subsection{The Gross-Pitaevskii equation}

We finally investigate domain formation in the Gross-Pitaevskii
equation. As has been remarked earlier, this formalism assumes
that the dynamics of the condensate is completely determined by
the precessional (and not relaxational) dynamics of the classical
order parameter. It is thus a formalism that on the one hand is
valid strictly at zero temperature, but on the other hand ignores
quantum fluctuations. The initial state chosen for models A and F
considered earlier does not evolve in this formalism and hence we
choose a slightly different initial state as mentioned in section
VII with 90$\%$ of the condensate density in the 0 state and 5$\%$
each in the +1 and -1 states. The precessional nature of the GP
equation implies that there is never any true relaxation to a
state with only domains of +1 and -1 and the amplitudes of these
two components oscillate together, $\pi/2$ out of phase with the
amplitude of the 0 component. The dynamical critical exponent $z$
in these simulations is extracted by looking at the time interval
when the amplitudes of the +1 and -1 components are growing with
time and that of the 0 component falling. It is important that a
sizeable window be identified within this interval where the
domain size $L(t)$ is indeed growing as $L(t) \propto t^{1/z}$.
The oscillatory nature of the dynamics ensures that in this case,
both the magnetization and condensate density are conserved. The
former is manifested in the fact the +1 and -1 components always
have the same amplitude and the latter in the fact that these two
components are always $\pi/2$ out of phase with the 0 component.

\begin{figure}[h!]
\epsfxsize=4in \centerline{\epsfbox{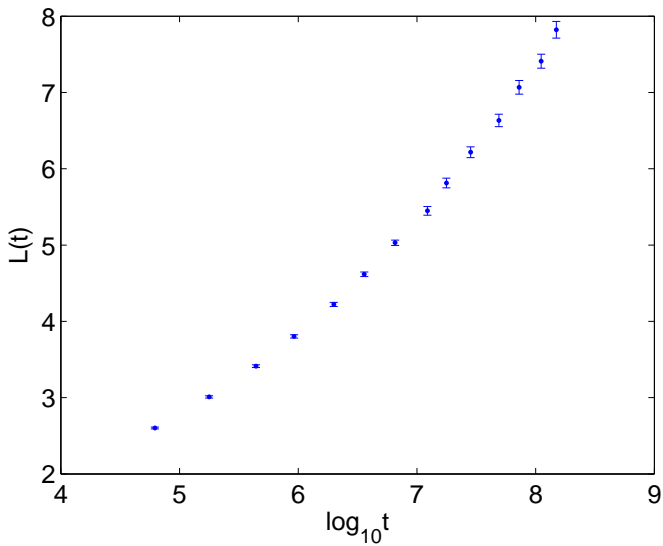}} \caption{$L(t)$ as a
function of $\log_{10}t$ in the GP equation with the parameter set
${\mathcal R}$. The data is from a time interval during which the
amplitudes of the $\pm 1$ components are increasing with time.}
\label{gpd}
\end{figure}

\begin{figure}[h!]
\epsfxsize=4in \centerline{\epsfbox{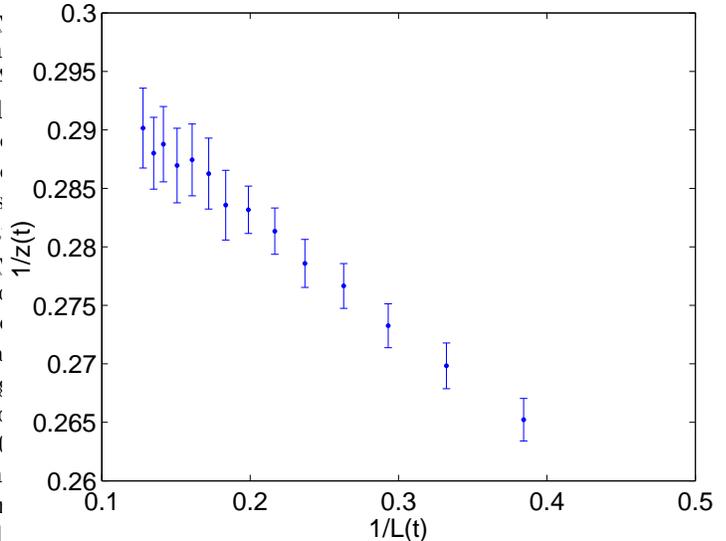}} \caption{$1/z(t)$
as a function of $1/L(t)$  in the GP equation with the parameter
set ${\mathcal R}$. The exponent $z(t)$ as a function of $t$ and
thus increasing $L(t)$ seems to drift towards the value $z=3$ like
with models A and F with conserved magnetization densities.}
\label{gpdrift}
\end{figure}

The domain size here is obtained only using Eqn.~\ref{domcalc}
since the presence of bubbles of the 0 state renders the method of
measuring the domain boundaries directly unreliable. The data for
the domain size $L(t)$ as a function of $t$ is shown in
Fig.~\ref{gpd}. The data as in the previous two cases has been
obtained over a range of about four decades. The dynamic critical
exponent $z(t)$ as extracted is shown as a function of $1/L(t)$.
Once again, there appears to be a drift towards the value $z=3$ at
infinite time, although in this case, it appears (to within the
noise) that the drift is more consistent with Eqn.~\ref{domdrift}
(i.e. linear in $1/L(t)$) than for models A and F.

It appears that the GP model gives a different dynamic critical
exponent $z=3$ from models A and F ($z=2$), unless magnetization
is conserved explicitly in the latter. This should be compared and
contrasted with the case of spinless bosons, where it has been
hypothesized that model F and the GP equation give the same
dynamic critical exponent for phase ordering~\cite{bosedyn}.
Further, this exponent was numerically found to be equal to 1
(different from model A, which has $z=2$) from numerical studies
of the GP equation in this case. The situation we analyze is
different from the case of spinless bosons in that we study the
magnetization. The second sound mode of model F arises from total
energy and number conservation between the condensate and the
``normal state''. The GP equation too conserves both these
quantities. However, for bosons with spin, the GP equation also
conserves magnetization, which is not present in model F unless
included by hand. Thus the value of $z$ for the GP equation is
different from that of model F and agreement is obtained only when
magnetization is explicitly conserved in the latter. It is
interesting to note however that the value of $z$ obtained from
the GP equation is larger than from model A in our study whereas
for spinless bosons it is smaller.

\section{Conclusions}

To conclude, we have studied the statics and dynamics of spin-1
condensates at finite temperature in the presence of a magnetic
field. We have obtained a ground state phase diagram for this
system and have focussed on the phase that is most amenable to a
numerical study of magnetic domain formation in 2D, namely the
Ferromagnetic out-of-plane phase and have numerically determined
the nature and order of the superfluid and magnetic phase
transitions. We have argued that the ``correct'' dynamical model
for spinor condensates at finite field and temperature is model F
in the Halpering and Hohenberg classification and have
demonstrated that this model contains all the modes in the
standard GP equation. We have numerically studied magnetic domain
formation in the GP model and models A and F with and without
magnetization conservation, and have found that it is the only
when magnetization is explicitly conserved in models A and F that
the dynamic critical exponenent $z$ obtained from the GP equations
agrees with the $z$ obtained from them. In the absence of this
conservation, models A and F yield $z=2$.

While we have focussed only on one part of the phase diagram of
spinor condensates in this study to highlight the difference
between various dynamical models, similar studies can be performed
in the other parts of the phase diagram as well. This will be
reported elsewhere. It is our belief that model F is fundamentally
a more complete dynamical model to describe spinor condensates
than the GP model.  In addition to studies of the dynamics far away from the critical point,
as presented here, this dynamical model could be used to
obtain dynamical critical exponents, for comparison to dynamical experiments
near the static phase transition of the spinor condensate.

\appendix
\section{Relation of the model F parameters to measurable quantities}

As remarked earlier, Model F has many more parameters than the GP
equation. Here we comment on how these parameters can be related
to experimentally measurable quantities. There are 9 important
real parameters, which are $\Gamma_1$, $\Gamma_2$, $g_0$,
$\lambda_0$, $\mu$, $c_0$, $c_2$, $C$, $\gamma$. Some of them are
directly measurable. For example, $\lambda_0$ is the thermal
conductivity, and $C$ is the specific heat. We now discuss how to
measure all the other parameters.

\subsection{$g_0$}

$g_0$ only exists in the dynamical equations and does not appear
in the static free energy. In the general formalism, at the
operator level, it appears in the Poisson bracket between $m$ and
$\psi_a$ as
\begin{equation}
[\psi_a^\dagger, m] = g_0\psi_a^\dagger
\end{equation}
This means if we create a particle through $\psi^\dagger$, the
expectation value of $m$ in the system increases by $g_0$. Since
$m$ is effectively the``heat'' in the system (this can be seen
from the physical meaning of $C$ and $\lambda$), $g_0$ is
effectively the ``heat'' per particle. Writing,
\begin{equation}
g_0 = T\sigma,
\end{equation}
where $\sigma$ is the entropy per particle, the value of $g_0$ can
now be obtained from measuring the specific heat
\begin{equation}
g_0 = T\int_0^TdT\frac{c}{T}.
\end{equation}
Here $c$ is the specific heat per particle.

\subsection{$c_0 + C\gamma^2$}

The reason we discuss $c_0$ and $\gamma$ together is that $c_0$
and $\gamma$ can only appear in the combination $c_0 + \gamma^2$
in all static quantities. This can be seen from mean field theory:
$m$ appears in a quadratic term $1/(2C)m^2$ and a linear term
$\gamma mn_0$: the expectation value of $m$ is thus $C\gamma$.
Plugging this value of $m$ into Eqn.~\ref{ginzburg2}, the $m^4$
interaction term becomes $c_0 + C\gamma^2$, which we denote as
$c^\prime$. How do we measure $c^\prime$? At low temperature,
almost all the atoms are in the condensate. All the terms in
Eqn.~\ref{ginzburg2} are proportional to the density of the
condensate, except the $c^\prime$ term which is proportional to
the square of the density. This term will therefore contribute to
the compressibility $\kappa$ of the condensate, where
\begin{equation}
\kappa^{-1} = -V\frac{dP}{dV} = V\frac{d^2E}{dV^2} = 2c^\prime
n^2_0
\end{equation}
When the temperature is low enough, the dominant contribution to
the compressibility will be from the condensate. By measuring the
compressibility, we can measure the parameter $c^\prime$.

\subsection{$\mu$}

Knowing the value of $c^\prime$, the value of $\mu$ is quite
straightforward to measure. It can be related to the density of
atoms within the condensate, using the relation
\begin{equation}
n_0 = \sqrt{\frac{\mu}{2c^\prime}}
\end{equation}

\subsection{$c_2$}

$c_2$ is a static parameter and should be measurable from static
properties, for example, the spin susceptibility. The terms in the
free energy involving the magnetization can be rewritten as
$c_2M_z^2 + hM_z$, where $h$ is the magnetic field along $z$. The
spin susceptibility is approximately $h/c_2$, from which we can
deduce $c_2$. In most practical cases at low temperature, the
value of $c_2$ is not very different from that obtained from the
atomic $s$ wave scattering rates.

\subsection{$\Gamma_2$}

The mode $\phi_{+1} + \phi_{-1}$ will have an oscillatory
component and also a part that is decaying. The value of the
oscillation frequency can be shown to be
$(1+\sqrt{5})\Gamma_2c_2n_0$. If we know the value of $c_2$ from
static experiments, we can obtain the value of $\Gamma_2$.

\subsection{$c_0$ and $\gamma$}

We have shown that $c_0+C\gamma^2$ can be determined by measuring
$c^\prime$. To disentangle the two quantities, we look at the
second sound velocity. The equation for the second sound velocity
is
\begin{equation}
c_s = \sqrt{\frac{g_0^2n_0}{C} + 2\Gamma_2^2(c^\prime +
\frac{g_0\gamma}{\Gamma_2})n_0}
\end{equation}
If we know $c_s$, the only unknown variable is $\gamma$. Having
obtained $\gamma $ from this equation, we can calculate $c_0$,
from the known value of $c^\prime$.

The authors wish to acknowledge conversations with D. A. Huse, A.
Lamacraft, S. Leslie, D. Podolsky, L. Sadler, D. M. Stamper-Kurn,
M. Vengalattore and A. Vishwanath, and support from DOE (S. M.),
NSF DMR-0238760 (C. X. and J. E. M.), and the IBM SUR program.

\bibliography{bigbib}
\end{document}